\documentclass[12pt,reqno,a4paper]{amsart}    %A4-format
\usepackage[totalwidth=450pt, totalheight=717pt]{geometry}
\usepackage{amsmath, amssymb}%, theorem}
\usepackage{amsthm}

\usepackage{graphicx}                      % for graphics
\newcommand{\color}[2][{}]{}         % figure-x.pstex_t file contains this
\usepackage{bbm}         % blackboard (see: /usr/share/texmf/mytex/bbm.dvi)

\usepackage{mathrsfs}    % use ``real'' caligraphic letters for \mathcal
\renewcommand\mathcal\mathscr

\numberwithin{equation}{section}

 \theoremstyle{plain}            % body italics
 \newtheorem{theorem}{Theorem}[section]

 \newtheorem{lemma}[theorem]{Lemma}
 \newtheorem{corollary}[theorem]{Corollary}

 \theoremstyle{definition}       % body roman

 \theoremstyle{remark}
 \newtheorem{remark}[theorem]{Remark}
 
\newcommand{\Sec}[1]{Section~\ref{sec:#1}}

\newcommand{\Fig}[1]{Figure~\ref{fig:#1}}

\newcommand{\Thm}[1]{Theorem~\ref{thm:#1}}

\newcommand{\ThmS}[2]{Theorems~\ref{thm:#1}--\ref{thm:#2}}

\newcommand{\Lem}[1]{Lemma~\ref{lem:#1}}
\newcommand{\Lems}[2]{Lemmata~\ref{lem:#1} and~\ref{lem:#2}}

\newcommand{\Cor}[1]{Corollary~\ref{cor:#1}}

\newcommand{\Rem}[1]{Remark~\ref{rem:#1}}

\newcommand{\dd}    {\, \mathrm d}    % not optimal: no \, if at beginning

\DeclareMathOperator{\dom}    {dom}
\DeclareMathOperator{\id}     {id}  % identity map
\DeclareMathOperator{\vol}    {vol}
\newcommand{\specsymb} {\sigma} % symbol for spectrum

\newcommand{\spec}[2][{}]   {\specsymb_{\mathrm{#1}}(#2)}

\newcommand{\disspec}[1]{\spec[disc]{#1}}

 \newcommand{\Err}{\mathrm O}
\def\Xint#1{\mathchoice
   {\XXint\displaystyle\textstyle{#1}}%
   {\XXint\textstyle\scriptstyle{#1}}%
   {\XXint\scriptstyle\scriptscriptstyle{#1}}%
   {\XXint\scriptscriptstyle\scriptscriptstyle{#1}}%
   \!\int}
\def\XXint#1#2#3{{\setbox0=\hbox{$#1{#2#3}{\int}$}
     \vcenter{\hbox{$#2#3$}}\kern-.5\wd0}}
\def\XXsum#1#2#3{{\setbox0=\hbox{$#1{#2#3}{\int}$}
     \vcenter{\hbox{$#2#3$}}\kern-.60\wd0}}

\newcommand{\dashint}{\Xint-}   % \int with -
\newcommand{\avint}{{\textstyle\dashint}}   % average sum
\newcommand{\R}{\mathbb{R}} % symbol for real numbers
\newcommand{\C}{\mathbb{C}} % symbol for complex numbers
\newcommand{\eps}{\varepsilon} % shortcut
\renewcommand{\phi}{\varphi}   % shortcut
\newcommand{\e}{\mathrm e}  %Euler number
\newcommand{\im}{\mathrm i} % complex unit

\newcommand{\wt}{\widetilde}           % shortcut
\newcommand {\qf}[1]{\mathfrak{#1}}    % font for quadratic forms

\newcommand{\HS}{\mathcal H}           % symbol for Hilbert space

\newcommand{\Sobsymb} {\mathsf H}      % symbol for Sobolev space
\newcommand{\Lsymb}    {\mathsf L}     % symbol for int L-spaces
\newcommand{\Lsqr}[2][{}]{\Lsymb_2^{#1}({#2})} % L_2(#1)-spaces
\newcommand{\Sob}[2][1]{\Sobsymb^{#1}({#2})}         % Sobolev space
\newcommand{\Sobx}[3][1]{\Sobsymb_{{#2}}^{#1}({#3})} % Sobolev space
\newcommand{\abs}[2][{}]{\lvert{#2}\rvert_{{#1}}}    % abs value
\newcommand{\abssqr}[2][{}]{\lvert{#2}\rvert^2_{#1}} % abs squared
\newcommand{\bigabs}[2][{}]{\bigl\lvert{#2}\bigr\rvert_{#1}}     % abs
\newcommand{\bigabssqr}[2][{}]{\bigl\lvert{#2}\bigr\rvert^2_{#1}}% abs squared
\newcommand{\Bigabssqr}[2][{}]{\Bigl\lvert{#2}\Bigr\rvert^2_{#1}}% abs squared
\newcommand{\norm}[2][{}]{\|{#2}\|_{{#1}}}    % norm
\newcommand{\normsqr}[2][{}]{\|{#2}\|^2_{#1}} % norm squared
\newcommand{\bignorm}[2][{}]{\bigl\|{#2}\bigr\|_{#1}}     % norm
\newcommand{\bignormsqr}[2][{}]{\bigl\|{#2}\bigr\|^2_{#1}}% norm squared
\newcommand{\iprod}[3][{}]{\langle{#2},{#3}\rangle_{#1}}  % inner product
\newcommand{\bigiprod}[3][{}]{\bigl\langle{#2},{#3}\bigr\rangle_{#1}}

\newcommand{\set}[2]{\{ \, #1 \, | \, #2 \, \} }      % set { #1 | #2 }
\newcommand{\bigset}[2]{\bigl\{ \, #1 \, \bigl|\bigr. \, #2 \, \bigr\} }
\newcommand{\Bigset}[2]{\Bigl\{ \, #1 \, \Bigl|\Bigr. \, #2 \, \Bigr\} }

\newcommand{\map}[3]{ #1 \colon #2 \longrightarrow #3}    % maps
\newcommand{\bd}  {\partial}                % symbol for boundary of a set
\newcommand{\clo}[1]{\overline{{#1}}} %            symbol for closure

\newcommand{\dcup}{\mathbin{\mathaccent\cdot\cup}}

\DeclareMathOperator*{\bigdcup}{\mathaccent\cdot{\bigcup}}

\newcommand{\restr}[1]{{\restriction}_{#1}} % symbol for map restriction

\newcommand{\conj}[1]{\overline {{#1}}}       % symbol for complex conjugation
\newcommand{\1}{\mathbf 1}                  % if bbm does'nt work
\renewcommand{\1}{\mathbbm 1}                    % blackboard 1
\newcommand{\und}{\quad\text{and}\quad}
\newcommand{\Und}{\qquad\text{and}\qquad}

\def\Xdashone{{-\mkern-12mu\1}}
\newcommand{\dashone}{\Xdashone}

\newcommand{\laplacian}[2][{}]{\Delta_{{#2}}^{{#1}}}
\newcommand{\lapl} [1][{}]{\Delta^{#1}} % symbol for Laplacian
\newcommand{\de} {\mathord{\mathrm d}} % exterior derivative

\newcommand{\ul}{\underline}

\newcommand{\Graph} X  % spacing?

\newcommand{\vxeps}{{\eps,v}}
\newcommand{\edeps}{{\eps,e}}
\newcommand{\Edeps}{{\eps,E}}
\newcommand{\vxed}{{v,e}}
\newcommand{\vxedeps}{{\eps,v,e}}
\newcommand{\cvol}{c_{\vol}}

\begin{document}
\title[Approximation of vertex couplings by Schr\"odinger operators on
manifolds]{Approximation of quantum graph vertex couplings by scaled
  Schr\"odinger operators on thin branched manifolds}

\author{Pavel Exner}
% \affiliation{%
%   Department of Theoretical Physics, NPI, Academy of Sciences,
%   25068 \v{R}e\v{z} near Prague, Czechia
% }
% \altaffiliation%
% [Also at ]{%
%   Doppler Institute, Czech Technical University, B\v{r}ehov\'{a}~7,
%   11519 Prague, Czechia
% }
\address{Department of Theoretical Physics, NPI, Academy of Sciences,
25068 \v{R}e\v{z} near Prague, and Doppler Institute, Czech
Technical University, B\v{r}ehov\'{a}~7, 11519 Prague, Czechia}
\email{exner@ujf.cas.cz}

\author{Olaf Post}      % remove \today in final version
% \affiliation{%
%          Institut f\"ur Mathematik,
%          Humboldt-Universit\"at zu Berlin,
%          Rudower Chaussee~25,
%          12489 Berlin,
%          Germany}
\address{Institut f\"ur Mathematik,
         Humboldt-Universit\"at zu Berlin,
         Rudower Chaussee~25,
         12489 Berlin,
         Germany}
\email{post@math.hu-berlin.de}
\date{\today \quad  \emph{File:} \texttt{\jobname.tex}}%, \currenttime h}
%\date{\today}

%------------------------------------------------------------
% Abstract.
%------------------------------------------------------------

\begin{abstract}
  We discuss approximations of vertex couplings of quantum graphs
  using families of thin branched manifolds.  We show that if a Neumann
  type Laplacian on such manifolds is amended by suitable potentials,
  the resulting Schr\"odinger operators can approximate non-trivial
  vertex couplings.  The latter include not only the $\delta$-couplings
  but also those with wavefunctions discontinuous at the vertex.  We
  work out the example of the symmetric $\delta'$-couplings and
  conjecture that the same method can be applied to all couplings
  invariant with respect to the time reversal.
\end{abstract}

% \pacs{Valid PACS appear here}% PACS, the Physics and Astronomy
%                              % Classification Scheme.

\maketitle

%----------------------------------------------------------------------
%
\section{Introduction}
\label{sec:intro}
%
%----------------------------------------------------------------------

The quantum graph models represent a simple and versatile tool to
study numerous physical phenomena.  The current state of art in this
field is described in the recent proceedings volume~\cite{ekkst:08} to
which we refer for an extensive bibliography.

One of the big questions in this area is the physical meaning of
quantum graph vertex coupling.  The general requirement of
self-adjointness admits boundary conditions containing a number of
parameters, and one would like to understand how to choose these
when a quantum graph model is applied to a specific physical
situation.  One natural idea is to approximate the graph in
question by a family of ``fat graphs'', i.e.\ tube-like manifolds
built around the graph ``skeleton'', equipped with a suitable
second-order differential operator.  Such systems have no ad hoc
parameters and one can try to find what boundary condition arise
when the manifold is squeezed to the graph.

The question is by no means easy and the answer depends on the type of
the operator chosen.  If it is the Laplacian with Dirichlet boundary
conditions one has to employ an energy renormalisation because the
spectral threshold given by the lowest transverse eigenvalue blows up
to infinity as the tube diameter tends to zero.  If one chooses the
reference point between the thresholds, the limiting boundary
conditions are determined by the scattering on the respective ``fat
star'' manifold \cite{molchanov-vainberg:07}.  If, on the other hand,
the threshold energy is subtracted, the limit gives generically a
decoupled graph, i.e.\ the family of edges with Dirichlet conditions at
their endpoints~\cite{post:05, molchanov-vainberg:07,
  dell-antonio-tenuta:06}.  One can nevertheless get a non-trivial
coupling in the limit if the tube network exhibits a threshold
resonance~\cite{grieser:08, acf:07}, and moreover, using a more
involved limiting process one can get also boundary conditions with
richer spectral properties~\cite{cacciapuoti-exner:07}.

The case when the fat graph supports a Laplacian of Neumann type is
better understood and the limit of all types of spectra as well as of
resonances has been worked out~\cite{freidlin-wentzell:93,
  rubinstein-schatzman:01,kuchment-zeng:01,exner-post:05,
  exner-post:07,grieser:08, exner-post:08a}.   Moreover, convergence of
resolvents etc.\ has been shown
in~\cite{saito:00,post:06,exner-post:07}.   Of course, no energy
renormalisation is needed in this case.  On the other hand, the limit
yields only the simplest boundary conditions called free or Kirchhoff.

The aim of this paper is to show that one can do better in the Neumann
case if the Laplacian is replaced by suitable families of
Schr\"odinger operators with properly scaled potentials.  Such
approximations have been shown to work on graphs
themselves~\cite{exner:97b,enz:01}, here we are going to ``lift'' them
to the tube-like manifolds\footnote{Another approach to approximation of nontrivial vertex conditions was proposed recently in~\cite{pavlov:07,pavlov:pre07}}.   First we will show
that using potentials supported by the vertex regions of the manifold
with the ``natural'' scaling, as $\eps^{-1}$ where $\eps$ is the tube
radius parameter, we can get the so-called $\delta$-coupling, the
one-parameter family with the wavefunctions continuous everywhere,
including at the vertex.  This shows, in particular, that one cannot
achieve such an approximation in a purely geometric way, with a
curvature-induced potential of the type~\cite{dek:01}, because the
latter scales typically as $\eps^{-2}$.   As main result in this case,
we show the convergence of the spectra and the resolvents (cf.\
\ThmS{closeness}{non-compact}).

On the other hand, the $\delta$-coupling is only a small part in the
set of all admissible couplings; in a vertex joining $n$ edges the
boundary conditions contain $n^2$ parameters.  Here we use the seminal
idea of Cheon and Shigehara~\cite{cheon-shigehara:98} applied to the
graph case in~\cite{cheon-exner:04} and generalised
in~\cite{exner-turek:05,exner-turek:07}.  For simplicity we will work
out in this paper the example of the so-called symmetric
$\delta'$-coupling, in short $\delta'_{\mathrm s}$, a one-parameter
family which is a counterpart of $\delta$, by using the result
of~\cite{cheon-exner:04} and ``lifting'' it to the manifold.  We show
that such a coupling is approximated with a potential in the vertex
region together with potentials at the edges with compact supports
approaching the vertex, all properly scaled, cf.\ \Thm{res.delta'}.
The speed with which the potentials are ``coming together'' must be
slower than the squeezing; the rate between the two we obtain is
surely not optimal.

We have no doubts that in the same way one can lift to the
manifolds the more general limiting procedure devised
in~\cite{exner-turek:07} which gives rise to a ${n+1 \choose
2}$-parameter family of boundary conditions, namely those which
are invariant with respect to the time reversal.  Such an extension
would be technically demanding, however, and in order not to
burden this paper with a complicated notation and voluminous
estimations we postpone it to a later work.

Let us survey the contents of the paper.  In the next section we
define the graph and manifold models and provide necessary estimates.
In \Sec{delta} we prove the convergence in the $\delta$-coupling case.
For simplicity we restrict ourselves to a star-shaped graph with a
single vertex; the approximation bears a local character and extends
easily to more complex graphs.  Finally in Section 4 we extend the
result to the $\delta'_{\mathrm s}$-coupling case and comment on the
applicability of the method to more general couplings.

%----------------------------------------------------------------------
%
\section{The graph and manifold models}
\label{sec:model}
%
%----------------------------------------------------------------------
%----------------------------------------------------------------------
\subsection{Graph model}
\label{sec:graph}
%----------------------------------------------------------------------
Let us start with a simple example of a star-shaped metric graph
$G=I_v$ having only one vertex $v$ and $\deg v$ adjacent edges $E=E_v$
of length $\ell_e \in (0,\infty]$, so we can think of
$E=\{1,\dots,\deg v\}$.  We identify the (metric) edge $e$ with the
interval $I_e:=(0,\ell_e)$ oriented in such a way that $0$ corresponds
to the vertex $v$.  Moreover, the metric graph $G=I_v$ is given by the
abstract space $I_v := \bigdcup_e \clo I_e / \sim$ where $\dcup$
denotes the disjoint union, and where the equivalence relation $\sim$
identifies the points $0 \in \clo I_e$ with the vertex $v$.  The basic
Hilbert space is $\Lsqr G:= \bigoplus_{e \in E} \Lsqr {I_e}$ with norm
given by
% ------------- %
\begin{equation*}
  \normsqr f = \normsqr[G] f
  = \sum_{e \in E} \int_0^{\ell_e} \abssqr {f(s)} \dd s.
\end{equation*}
% ------------- %
The decoupled Sobolev space of order $k$ is defined as
% ------------- %
\begin{equation*}
  \Sobx[k] \max G := \bigoplus_{e \in E} \Sob[k] {I_e}
\end{equation*}
% ------------- %
together with its natural norm.   Let $\ul p=\{p_e\}_e$ be a vector
consisting of the weights $p_e>0$ for $e \in E$.  The Sobolev space
associated to the weight $\ul p$ is given by
% ------------- %
\begin{equation}
  \label{eq:sob1}
  \Sobx {\ul p} G
     := \bigset{f \in \Sobx \max G} {\ul f \in \C \ul p},
\end{equation}
% ------------- %
where $\ul f := \{f_e(0)\}_e \in \C^{\deg v}$ is the evaluation
vector of $f$ at the vertex $v$ and $\mathbb{C}{\ul p}$ is the
complex span of $\ul p$.  We use the notation
% ------------- %
\begin{equation}
  \label{eq:eval.f}
  \ul f = f(v) \ul p, \qquad\text{i.e.,} \qquad
  f_e = f(v) p_e
\end{equation}
% ------------- %
for all $e \in E$.  In particular, if $\ul p=(1,\dots,1)$, we
arrive at the \emph{continuous} Sobolev space $\Sob G := \Sobx
{\ul p} G$.  The standard Sobolev trace estimate
% ------------- %
\begin{equation}
  \label{eq:sob.tr}
  \abssqr {g(0)}
  \le a \normsqr[(0,\ell)]{g'} + \frac 2 a \normsqr[(0,\ell)] g
\end{equation}
% ------------- %
for $g \in \Sob{0,\ell}$ and $0 < a \le \ell$ ensures that $\Sobx
{\ul
  p} G$ is a closed subspace of $\Sobx \max G$, and therefore itself a
Hilbert space.  A simple consequence is the following claim.
% ------------- %
\begin{lemma}
  \label{lem:sob.tr}
  We have
% ------------- %
  \begin{equation*}
    \abssqr{f(v)}
     \le \abs{\ul p}^{-2}
      \Bigl( a \normsqr[G] {f'}
            + \frac 2 a \normsqr[G] f
      \Bigl)
  \end{equation*}
% ------------- %
  for $f \in \Sobx {\ul p} G$ and $0 < a \le \ell_- := \min_{e \in E}
  \ell_e$, the minimal length at the vertex $v$.
\end{lemma}
% ------------- %

We define various Laplacians on the metric graph via their
quadratic forms.  Let us start with the (weighted) \emph{free}
Laplacian $\lapl_G$ defined via the quadratic form $\qf d=\qf d_G$
given by
% ------------- %
\begin{equation*}
  \qf d (f) := \normsqr[G] {f'} = \sum_e \normsqr[I_e]{f_e'} \Und
  \dom \qf d := \Sobx{\ul p} G
\end{equation*}
% ------------- %
for a fixed ${\ul p}\:$ (the forms and the corresponding operators
should be labelled by the weight ${\ul p}$, of course, but we drop
the index, in particular, because we are most interested in the
case $\ul p=(1,\dots,1)$).  Note that $\qf d$ is a closed form
since the norm associated to the quadratic form $\qf d$ is
precisely the Sobolev norm given by $\normsqr[\Sob G]
f=\normsqr[G] {f'} + \normsqr[G] f$.

The Laplacian with \emph{$\delta$-coupling of strength $q$} is
defined via the quadratic form $\qf h=\qf h_{(G,q)}$ given by
% ------------- %
\begin{equation}
  \label{eq:def.qf.delta}
  \qf h(f) := \normsqr[G] {f'} + q(v) \abssqr{f(v)} \Und
  \dom \qf h := \Sobx {\ul p} G.
\end{equation}
% ------------- %
The $\delta$-coupling is a ``small'' perturbation of the free
Laplacian, namely we have:
% ------------- %
\begin{lemma}
  \label{lem:delta.pert}
  The form $\qf h_{(G,q)}$ is relatively form-bounded with respect to
  the free form $\qf d_G$ with relative bound zero, i.e., for any
  $\eta>0$ there exists $C_\eta>0$ such that
% ------------- %
  \begin{equation*}
    \abs{\qf h(f) - \qf d(f)}
    = \abs{q(v)}\abssqr{f(v)}
    \le \eta \, \qf d(f) + C_\eta \normsqr[G] f.
  \end{equation*}
% ------------- %
\end{lemma}
% ------------- %
\begin{proof}
  It is again a simple consequence of~\Lem{sob.tr}.  Since we
  need the precise behaviour of the constant $C_\eta$, we give a short
  proof here.  From \Lem{sob.tr} we conclude that
% ------------- %
  \begin{equation*}
    \abs{\qf h(f) - \qf d (f)}
    \le \abs{q(v)} \abs{\ul p}^{-2}
          \Bigl(a \qf d(f) + \frac 2 a \normsqr[G] f \Bigr).
  \end{equation*}
% ------------- %
  for any $0 < a \le \ell_-$.  Set $a := \min \{\eta \abssqr{\ul
    p}/\abs{q(v)}, \ell_-\}$ and
% ------------- %
  \begin{equation*}
    C_\eta := 2\max \Bigl\{ \frac{\abssqr{q(v)}}{\eta \abs{\ul p}^4},
                 \frac{\abs{q(v)}}{\ell_- \abssqr{\ul p} } \Bigr\},
  \end{equation*}
% ------------- %
  then the desired estimate follows.
\end{proof}
% ------------- %
One can see that the norms associated to $\qf h$ and $\qf d$ are
equivalent and, in particular, setting $\eta=1/2$ in the above
estimate yields we get:
% ------------- %
\begin{corollary}
  \label{cor:delta.pert}
  The quadratic form $\qf h$ is closed and obeys the estimate
% ------------- %
  \begin{equation*}
    \qf d (f) \le 2 (\qf h (f) + C_{1/2} \normsqr[G] f).
  \end{equation*}
% ------------- %
\end{corollary}
% ------------- %

The operator $H=H_{(G,q)}$ associated to $\qf h$ acts as $(H
f)_e=-f_e''$ on each edge and satisfies the conditions
% ------------- %
\begin{equation}
  \label{eq:vx.cond}
    \frac{f_{e_1}(0)}{p_{e_1}} = \frac{f_{e_2}(0)}{p_{e_2}}=:f(v)
    \Und
    \sum_e p_e f_e'(0) = q(v) f(v)
\end{equation}
% ------------- %
for any pair $(e_1,e_2)$ of edges meeting at the vertex $v$.  We
use the formal notation
% ------------- %
\begin{equation}
  \label{eq:delta}
  H= H_{(G,q)} = \lapl_G + q(v) \delta_v\,;
\end{equation}
% ------------- %
note that $\lapl_G$ is a non-negative operator by definition.  In
order to compare the ``free'' quadratic form with the graph norm
of $H$ we need the following estimate:
\begin{lemma}
  \label{lem:res.est}
  We have
  \begin{equation*}
    \normsqr[\Sob G] f
    = \qf d(f) + \normsqr[G] f
    \le 2 \max \{ C_{1/2},\sqrt 2\} \normsqr[G]{(H-\im)f}
  \end{equation*}
  for $f \in \dom H \subset \dom \qf h = \Sobx {\ul p} G$.
\end{lemma}
\begin{proof}
  Using the estimate of \Cor{delta.pert}, we obtain
  \begin{equation*}
    \qf d(f) + \normsqr f
    \le 2 \bigl(\qf h(f) + (C_{1/2}+1)\normsqr f \bigr)
    \le 2 \bigabs{\qf h(f) + \normsqr f}
               +     2 C_{1/2} \normsqr f.
  \end{equation*}
  Moreover, the first term can be estimated as
  \begin{multline*}
    \bigabssqr{\qf h(f) + \normsqr f}
    \le 2 \bigl(\qf h(f)^2 + \norm f^4\bigr)
    = 2 \bigabssqr{\qf h(f) - \im \normsqr f}
    = 2 \bigabssqr{\iprod f {(H-\im)f}}\\
    \le 2 \normsqr f \normsqr{(H-\im)f}.
  \end{multline*}
  Finally, we apply the estimate $\norm f \le \norm{(H-\im)f}$ to
  obtain the result.
\end{proof}
% ------------- %
\begin{remark}
  Note that we have not said anything about the boundary conditions at
  the free ends of the edges of finite length if there are any.  As we
  employ the Sobolev space $\Sobx {\ul p} G$ for the domain, we implicitly
  introduce Neumann conditions for the operator, $f'_e(\ell_e)=0$.  However,
  one can choose any other condition at the free ends, or to construct more
  complicated graphs by putting the star graphs together.
\end{remark}
% ------------- %

%----------------------------------------------------------------------
\subsection{Manifold model of the ``fat'' graph}
\label{sec:mfd}
%----------------------------------------------------------------------
Let us now define the other element of the approximation we are
going to construct.  For a given $\eps \in (0, \eps_0$] we
associate a $d$-dimensional manifold $X_\eps$ to the graph $G$ in
the following way.  To the edge $e \in E$ and the vertex $v$ we
ascribe the Riemannian manifolds
% ------------- %
\begin{equation}
  \label{eq:mfd.ed}
  X_\edeps := I_e \times \eps Y_e \Und
  X_\vxeps := \eps X_v,
\end{equation}
% ------------- %
respectively, where $\eps Y_e$ is a manifold $Y_e$ equipped with
metric $h_\edeps:=\eps^2 h_e$ and $\eps X_\vxeps$ carries the metric
$g_\vxeps=\eps^2 g_v\,$; here $h_e$ and $g_v$ are $\eps$-independent
metrics on $Y_e$ and $X_v$, respectively.   We identify the boundary
component $\bd_e X_\vxeps=\eps \bd_e X_v$ of $\bd X_\vxeps=\eps \bd
X_v$ with $\bd_v X_\edeps=\eps \bd_v X_e = \{0\} \times \eps Y_e$ and
call the resulting manifold $X_\eps$.   We refer to the unscaled
manifold (which conventionally means $\eps=1$) as to $X$.  In
particular, the manifold $X$ consists of the number $\deg v$ of
straight cylinders\footnote{The straightness here refers to the
  intrinsic geometry only.  We do not assume in general that the
  manifolds $X_\eps$ are embedded, for instance, into a Euclidean
  space, see also \Rem{long.err}.} with cross-section $Y_e$ and a
vertex neighbourhood manifold $X_v$ containing the boundary components
$Y:= \bigdcup_e Y_e$.  Without loss of generality, we may assume that
each cross-section $Y_e$ is connected, otherwise we replace the edge
$e$ by as many edges as is the number of connected components.  We
denote the boundary component of $\bd X_v$ at the edge $e$ by $\bd_e
X_v$ and the boundary component of $\bd X_e$ at the vertex $v$ by
$\bd_v X_e= \{0\} \times Y_e$.  Note that these two boundaries are
identified in the entire manifold $X$.   Similarly, we denote by $\bd_e
X_\vxeps=\eps \bd_e X_v$ and $\bd_v X_\edeps =\eps \bd_v X_e$ their
scaled versions.  For convenience, we will always use the
$\eps$-independent coordinates $(s,y) \in X_e=I_e \times Y_e$ and $x
\in X_v$, so that the radius-type parameter $\eps$ only enters via the
Riemannian metrics.

The entire manifold $X_\eps$ may or may not have a boundary $\bd
X_\eps$, depending on whether there is at least one finite edge
length $\ell_e<\infty$ or one ``transverse'' manifold $Y_e$ has a
non-empty boundary.  In such a situation, we assume that $X_\eps$ is
open in $\clo X_\eps = X_\eps \dcup \bd X_\eps$.  A particular case
is represented by embedded manifolds which deserve a comment:
% ------------- %
\begin{remark}
  \label{rem:long.err}
  Note that the above setting contains the case of the
  $\eps$-neighbourhood of an embedded graph $G \subset \R^2$, but
  only  up to a longitudinal error of order of $\eps$.   The manifold
  $X_\eps$ itself does \emph{not} form an $\eps$-neighbourhood of a
  metric graph embedded in some ambient space, since the vertex
  neighbourhoods cannot be fixed in the ambient space unless one allows
  slightly shortened edge neighbourhoods.   Nevertheless, introducing
  $\eps$-independent coordinates also in the longitudinal direction
  simplifies the comparison of the Laplacian on the metric graph and
  the manifold, and the error made is of order of $\Err(\eps)$, as we
  will see in \Lem{long.err} for a single edge.
\end{remark}
% ------------- %
\noindent The basic Hilbert space of the manifold model is
\begin{equation}
  \label{eq:lsqr.xeps}
  \Lsqr {X_\eps}
     = \bigoplus_e \bigl(\Lsqr {I_e} \otimes \Lsqr {\eps Y_e}\bigr)
        \oplus \Lsqr {\eps X_v}
\end{equation}
with the norm given by
% ------------- %
\begin{equation*}
  \normsqr[X_\eps] u
  = \sum_{e \in E} \eps^{d-1} \int_{X_e} \abssqr u \dd y_e \dd s
   + \eps^d \int_{X_v} \abssqr u \dd x_v
\end{equation*}
% ------------- %
where $\dd x_e=\dd y_e \dd s$ and $\dd x_v$ denote the Riemannian
volume measures associated to the (unscaled) manifolds $X_e=I_e
\times Y_e$ and $X_v$, respectively.  In the last formula we have
employed the appropriate scaling behaviour, $\dd x_\edeps =
\eps^{d-1} \dd y_e \dd s$ and $\dd x_\vxeps = \eps^d \dd x_v$.

Denote by $\Sob{X_\eps}$ the Sobolev space of order one, the
completion of the space of smooth functions with compact support
under the norm given by $\normsqr[\Sob {X_\eps}] u =
\normsqr[X_\eps] {\de u} + \normsqr[X_\eps] u$.  As in the case of
the metric graphs, we define the Laplacian $\lapl_{X_\eps}$ on
$X_\eps$ via its quadratic form
% ------------- %
\begin{equation}
  \label{eq:h.eps}
  \qf d_\eps (u)
  := \normsqr[X_\eps] {\de u}
  = \sum_{e \in E} \eps^{d-1} \int_{X_e}
          \Bigl(\abssqr {u'(s,y)}
                + \frac 1 {\eps^2} \abssqr[h_e]{\de_{Y_e} u}
          \Bigr) \dd y_e \dd s
   + \eps^{d-2} \int_{X_v} \abssqr[g_v] {\dd u} \dd x_v
\end{equation}
% ------------- %
where $u'$ denotes the \emph{longitudinal} derivative,
$u'=\partial_s u$, and ${\dd u}$ is the exterior derivative of
$u$.  As before, the form $\qf d_\eps$ is closed by definition.
Adding a potential, we define the Hamiltonian $H_\eps$ as the
operator associated with the form $\qf h_\eps = \qf
h_{(X_\eps,Q_\eps)}$ given by
% ------------- %
\begin{equation*}
  \qf h_\eps = \normsqr[X_\eps] {\de u} + \iprod[X_\eps] u {Q_\eps u}
\end{equation*}
   % ------------- %
where the potential $Q_\eps$ has support only in the (unscaled) vertex
neighbourhood $X_v$ and
   % ------------- %
\begin{equation}
  \label{eq:pot.eps}
  Q_\eps(x) = \frac 1 \eps Q(x)
\end{equation}
   % ------------- %
where $Q=Q_1$ is a fixed bounded and measurable function on $X_v$.
The reason for this particular scaling will become clear in the
proof of \Lem{closeness}.  Roughly speaking, it comes from the fact
that $\vol X_\vxeps$ is of order $\eps^d$, whereas the
$(d\!-\!1)$-dimensional transverse volume $\vol Y_\edeps$ is of
order $\eps^{d-1}$.  The operators $H_\eps$ and $\Delta_\eps$ are
associated to forms $\qf h_\eps$ and $\qf d_\eps$, respectively;
note that $\Delta_\eps = \lapl_{X_\eps} \ge 0$ is the usual
(Neumann) Laplacian on $X_\eps$.  As usual the Neumann boundary
condition occurs only in the operator domain if $\bd X_\eps \ne
\emptyset$.  We postpone for a moment the check that $H_\eps$ is
relatively form-bounded with respect to $\lapl_{X_\eps}$, see
\Lem{ham.pert} below.

Let us compare the two cylindrical neighbourhoods, $X_\edeps=I \times
\eps Y_e$ and $\wt X_\edeps =I_\eps \times \eps Y_e$, on edges of
length $\ell>0$ and $\ell_\eps=(1-\eps)\ell$, respectively.  The
result for the entire space $X_\eps$ then follows by combining the
estimates on the edges and the fact that the potential is only
supported on the vertex neighbourhoods.  The verification of the
conditions of $\delta$-closeness in the next lemma is straightforward;
for more details on $\delta$-closeness we refer
to~\cite[App.~A]{post:06} or~\cite{post:pre08a}.
%----------------------------------------------------------------------
\begin{lemma}
  \label{lem:long.err}
  Let $\HS_e:=\Lsqr{X_\edeps}$ and $\wt \HS_e:= \Lsqr{\wt X_\edeps}$.
  Moreover, define
  \begin{align*}
    \map {J_e&} {\HS_e} {\wt \HS_e}&
    (J_ef)(\wt s,y)&:=f((1-\eps)^{-1}\wt s,y),\\
    \map {J_e'&} {\wt \HS_e} {\HS_e}&
    (J_e'u)(s,y)&:=f((1-\eps)s,y).
  \end{align*}
  Then the quadratic forms $\qf d_\eps(f):=\normsqr[X_\edeps] f$ and
  $\wt{\qf d}_\eps(u):=\normsqr[\wt X_\edeps] u$ with $\dom \qf
  d_\eps = \Sob{X_\edeps}$ and $\dom \wt{\qf d}_\eps = \Sob{\wt
    X_\edeps}$ are $\delta_\eps$-close with $\delta_\eps=2
  \eps/(1-\eps)^{1/2}$; namely, we have $J_e'J_e=\id $, $J_e
  J'_e=\id$, $\norm {J_e} \le 1$, $\norm{J_e'} \le 1+\delta_\eps$,
  \begin{equation*}
    \norm{J_e' - J_e^*} \le \delta_\eps
        \Und
    \bigabs {\wt{\qf d}_\eps (J_e f, u) - \qf d_\eps(f, J'_e u)}
    \le \delta_\eps.
  \end{equation*}
  In particular, we get
  \begin{equation*}
    \norm{(\laplacian{\wt X_\edeps} + 1)^{-1}
        - J_e(\laplacian{X_\edeps} + 1)^{-1} J_e'}
  \le 2\delta_\eps=\Err(\eps).
  \end{equation*}
\end{lemma}
%----------------------------------------------------------------------
Before we check the closeness assumptions of~\cite[Appendix]{post:06}
in the next section, we need some more notation and estimates.  The
estimates are already provided in~\cite{exner-post:05,post:06}, but we
will also need a precise control of the edge length, when we
approximate the $\delta'_\mathrm{s}$-coupling by $\delta$-couplings in
\Sec{delta'} below.  Therefore, we present short proofs of the
estimate here.

We first introduce the following averaging operators
 % ------------- %
\begin{equation*}
  \avint_v u := \dashint_{X_v} u \dd x_v
           \Und
  \avint_e u(s) := \dashint_{Y_e} u(s,\cdot) \dd y_e
\end{equation*}
 % ------------- %
for $u \in \Lsqr {X_\eps}$, where we use the following symbols
\begin{equation*}
  \dashint_{X_v} u \dd x_v := \frac 1 {\vol_d X_v} \int_{X_v} u  \dd x_v
  \Und
  \dashint_{Y_e} \phi \dd y_e
         := \frac 1 {\vol_{d-1} Y_e} \int_{Y_e} \phi \dd y_e
\end{equation*}
denoting the \emph{normalised} integrals.  For brevity, we also
omit the measure and write $\avint_{\bd X_v} u\:$ etc.

In order to obtain the below Sobolev trace
estimate~\eqref{eq:sob.tr1}, we need a further decomposition of the
vertex neighbourhood $X_v$.  Recall that $X_v$ has $(\deg v)$-many
boundary components isometric to $Y_e$.  We assume that each such
boundary component has a collar neighbourhood $X_\vxed
=(0,\ell_e)\times Y_e$ of length $\ell_e$.  Note that the scaled
vertex neighbourhood $X_\vxeps=\eps X_v$ is of order $\eps$ in
\emph{all} directions, so that the scaled collar neighbourhoods
$X_\vxedeps := \eps X_\vxed$ are of length $\eps\ell_e$.  We can
always assume that such a decomposition exists, by possibly using a
different cut of the manifold $X$ into $X_v$ and $X_e$, the price
being an additional longitudinal error of order $\eps$ (see
\Lem{long.err}).  Similarly as in~\eqref{eq:sob.tr}, one can get the
following Sobolev trace estimates for the scaled manifolds:
\begin{gather}
  \label{eq:sob.tr1}
  \normsqr[\bd_e X_\vxeps] u
  \le \eps \wt a \normsqr[X_\vxedeps]{\de u}
     + \frac 2 {\eps \wt a} \normsqr[X_\vxedeps] u\\
  \label{eq:sob.tr2}
  \normsqr[\bd_v X_\edeps] u
  \le a \normsqr[X_\edeps]{u'}
     + \frac 2 a \normsqr[X_\edeps] u
\end{gather}
for $0 < a, \wt a \le \ell_e$ on the vertex and edge neighbourhood,
respectively, where $u'=\partial_s u$ denotes the longitudinal
derivative.  The unscaled versions are obtained, of course, by setting
$\eps=1$.  Moreover, by the Cauchy-Schwarz inequality we get
\begin{equation*}
  \vol Y_e \bigabssqr{\avint_e u(0)}
  \le \normsqr[\bd_e X_v] u
  =   \normsqr[\bd_v X_e] u.
\end{equation*}

In the following lemma we compare the averaging over the boundary of
$X_v$ with the averaging over the whole space $X_v$:
\begin{lemma}
  \label{lem:av.int}
  For $u \in \Sob {X_v}$, we have
  \begin{equation*}
    \vol \bd X_v \bigabssqr{\avint_{\bd X_v} u - \avint_v u}
    \le \sum_{e \in E}
    \vol \bd_e X_v \bigabssqr{\avint_{\bd_e X_v} u - \avint_v u}
    \le \Bigl( \wt a
             + \frac 2 {\wt a \lambda_2(v)}
        \Bigr) \normsqr[X_v] {\de u}
  \end{equation*}
  for $0 < \wt a \le \ell_- = \min_e \ell_e$, where $\lambda_2(v)$
  denotes the second (i.e., first non-zero) eigenvalue of the Neumann
  Laplacian on $X_v\,$; the latter is defined conventionally as the
  operator associated to the form $\qf d_v(u):= \normsqr[X_v]{\de u}$
  with the domain $\dom \qf d_v := \Sob {X_v}$.
\end{lemma}
\begin{proof}
  Using Cauchy-Schwarz twice and the estimate~\eqref{eq:sob.tr1} for
  each edge $e$ and $\eps=1$, we obtain
  \begin{equation}
  \label{eq:sob.tr.vx}
    \vol \bd X_v \bigabssqr{\avint_{\bd X_v} w}
    \le \sum_e \vol \bd_e X_v \bigabssqr{\avint_{\bd_e X_v} w}
    \le \normsqr[\bd X_v] u
    \le \wt a \normsqr[X_v]{\de w}
     + \frac 2 {\wt a} \normsqr[X_v] w
  \end{equation}
  for $0 < \wt a \le \ell_-$, using the fact that $\bigcup_e X_\vxed
  \subset X_v$.  We apply the above estimate to the function
  $w=P_vu:=u-\avint_v u$ and observe that
  \begin{equation}
    \label{eq:min-max}
    \normsqr[X_v] w \le \frac 1 {\lambda_2(v)} \normsqr[X_v]{\de w}
  \end{equation}
  as one can check using the fact that that $\de w = \de u$ and that
  $P_v$ is the projection onto the orthogonal complement of the first
  eigenfunction $\1_v \in \Lsqr {X_v}$.
\end{proof}

We also need an estimate over the vertex neighbourhood.  It will assure
that in the limit $\eps \to 0$, no family of normalised eigenfunctions
$(u_\eps)_\eps$ with eigenvalues lying in a bounded interval can
concentrate on $X_\vxeps$.
\begin{lemma}
  \label{lem:vx.est}
  We have
  \begin{equation*}
    \normsqr[X_\vxeps] u
    \le 4 \eps^2 \Bigl[ \frac 1 {\lambda_2(v)}
         + \cvol \Bigl( \wt a + \frac 2 {\wt a \lambda_2(v)}\Bigr)\Bigr]
          \normsqr[X_\vxeps] {\de u}
       + 4 \eps \cvol \Bigl[ a \normsqr[X_\Edeps]{u'}
                        + \frac 2 a \normsqr[X_\Edeps] u \Bigr]
  \end{equation*}
  for $0 < a,\wt a \le \ell_-= \min_e \ell_e$, where
  $\cvol:=\cvol(v)=\vol X_v/\vol\bd X_v$ and $X_\Edeps := \bigdcup_e
  X_\edeps$ denotes the union of all edge neighbourhoods.
\end{lemma}
\begin{proof}
  We start with the estimate
  \begin{equation*}
    \normsqr[X_\vxeps] u
    \le 2 \eps^d
       \Bigl( \bignormsqr[X_v]{u - \avint_v u}
            + \bignormsqr[X_v]{\avint_v u}
       \Bigr)
    \le 2 \eps^d
       \Bigl( \frac 2 {\lambda_2(v)} \normsqr[X_v]{\de u}
            +  \vol X_v \bigabssqr{\avint_v u}
       \Bigr)
  \end{equation*}
  using \Lem{av.int} and the fact that $\avint_v u$ is
  constant.  Moreover, the last term can be estimated by
  \begin{align*}
     \vol \bd X_v \bigabssqr{\avint_v u}
     &\le 2 \vol \bd X_v
          \Bigl( \bigabssqr{\avint_v u - \avint_{\bd X_v} u}
               + \bigabssqr{\avint_{\bd X_v} u}
          \Bigr) \\
     &\le 2
          \Bigl( \wt a + \frac 2 {\wt a \lambda_2(v)}
          \Bigr) \normsqr[X_v]{\de u}
               + \sum_e \vol \bd_e X_v
                      \bigabssqr{\avint_{\bd_e X_v} u}
  \end{align*}
  using~\eqref{eq:sob.tr.vx}.  Since $\bd_e X_v$ is isometric to $\bd_v
  X_e$, we can estimate the latter sum by
  \begin{equation*}
    \sum_e \vol \bd_v X_e \bigabssqr{\avint_{\bd_e X_v} u}
    \le \sum_e \normsqr[\bd_v X_e] u
    \le a \normsqr[X_E]{u'}
       + \frac 2 a \normsqr[X_E] u
  \end{equation*}
  due to~\eqref{eq:sob.tr2} for $\eps=1$, each edge $e$ and $0 < a \le
  \ell_-$.  Here, $X_E := X_{1,E}$ is the union of the unscaled edge
  neighbourhoods.  The desired estimate then follows from the scaling
  behaviour $\normsqr[X_\vxeps]{\de u}= \eps^{d-2} \normsqr[X_v] {\de
    u}$ and $\normsqr[X_\edeps] w = \eps^{d-1} \normsqr[X_e] w$ for
  $w=u$ or $w=u'$ (where $u'=\partial_s u$ denotes the longitudinal
  derivative).
\end{proof}

We are now able to prove the relative (form-)boundedness of the
Hamiltonian $H_\eps$ with respect to the Laplacian $\lapl_{X_\eps}$
for the indicated class of potentials.  It is again important here to
have a precise control of the constants $\eps_\eta$ and $\wt C_\eta$
in terms of the various parameters of our spaces.  This will be of
particular importance when we deal with the approximation of the
$\delta'_\mathrm{s}$-coupling by $\delta$-couplings with shrinking
spacing $a=\eps^\alpha$ in \Sec{delta'} below.
%----------------------------------------------------------------------
\begin{lemma}
  \label{lem:ham.pert}
  To a given $\eta \in (0,1)$ there exists $\eps_\eta>0$ such that the form $\qf
  h_\eps$ is relatively form-bounded with respect to the free form
  $\qf d_\eps$ with relative bound $\eta$ for all $\eps\in (0, \eps_\eta]$,
  in other words, there exists $\wt C_\eta>0$ such that
% ------------- %
  \begin{equation*}
    \abs{\qf h_\eps(u) - \qf d_\eps(u)}
    \le \eta \, \qf d_\eps(u) + \wt C_\eta \normsqr[X_\eps] u
  \end{equation*}
% ------------- %
  whenever $0 < \eps\le \eps_\eta$, where the constants $\eps_\eta$ and $\wt
  C_\eta$ are given by
  \begin{subequations}
    \label{eq:def.eps.eta}
    \begin{gather}
      \eps_\eta=\eps_\eta(\norm[\infty] Q,\ell_-) := \frac \eta {4
        \norm[\infty] Q} \Bigl[ \frac 1 {\lambda_2(v)} + \cvol \cdot
      \Bigl( \ell_- + \frac 2 {\ell_- \lambda_2(v)} \Bigr)
      \Bigr]^{-1},\\
      \wt C_\eta =\wt C_\eta(\norm[\infty] Q, \ell_-) := 8 \cvol
      \norm[\infty] Q \max \Bigl\{ \frac {4 \cvol \norm[\infty] Q}
      \eta, \frac 1{\ell_-} \Bigr\}.
    \end{gather}
  \end{subequations}
\end{lemma}
Note that $\eps_\eta=\Err(\ell_-)$ and $\wt C_\eta =
\Err(\ell_-^{-1})$ as $\ell_- \to 0$.
%----------------------------------------------------------------------
\begin{proof}
  The potential $Q_\eps=\eps^{-1}Q$ is by assumption supported on the vertex
  neighbourhood $X_v$, therefore we have
% ------------- %
  \begin{multline*}
    \abs{\qf h_\eps(f) - \qf d_\eps(f)}
    \le \frac {\norm[\infty] Q} \eps \normsqr[X_\vxeps] u\\
    \le 4 \norm[\infty] Q
      \Bigl\{
        \eps \Bigl[ \frac 1 {\lambda_2(v)}
         + \cvol \cdot \Bigl( \ell_-
                           + \frac 2 {\ell_- \lambda_2(v)}
                       \Bigr)
        \Bigr]  \normsqr[X_\vxeps]{\de u}
         + a \cvol \normsqr[X_\Edeps] {u'}
      \Big\}\\
      +\frac {8 \norm[\infty] Q \cvol} a \normsqr[X_\Edeps] u
  \end{multline*}
  using \Lem{vx.est}, for $0 < a \le \ell_-$ and $\wt a := \ell_-$.
  Choosing $a = \min \{ \ell_-, \eta(4\cvol \norm[\infty] Q)^{-1}\}$
  and $0 < \eps \le \eps_\eta$ with $\eps_\eta$ as above, we can estimate
  the quadratic form contributions by
  \begin{equation*}
    \eta \bigl(\normsqr[X_\vxeps] {\de u} +
               \normsqr[X_\Edeps] {u'}
         \bigr)
    \le \eta \normsqr[X_\eps] {\de u}.
  \end{equation*}
  The expression for $\wt C_\eta$ then follows by evaluating the
  constant in front of the remaining norm.
\end{proof}
% ------------- %

We need to estimate the ``free'' quadratic form against the form
associated with the Hamiltonian:
% ------------- %
\begin{corollary}
  \label{cor:ham.pert}
  The quadratic form $\qf h_\eps$ is closed.  Moreover, setting
  $\eta=1/2$, we get the estimate
  % ------------- %
  \begin{equation*}
    \qf d_\eps (u)
    \le 2  \bigl( \qf h_\eps (u)
             + \wt C_{1/2} \normsqr[X_\eps] u\bigr)
  \end{equation*}
  % ------------- %
  which holds provided $0 < \eps \le \eps_{1/2}$.
\end{corollary}
% ------------- %
As in \Lem{res.est}, we can prove the following estimate in order to
compare the ``free'' quadratic form with the graph norm of $H_\eps$:
\begin{lemma}
  \label{lem:res.est.mfd}
  We have
  \begin{equation*}
    \normsqr[\Sob {X_\eps}] u
    = \qf d_\eps(u) + \normsqr[X_\eps] u
    \le 2 \max \{ \wt C_{1/2},\sqrt 2\}
             \normsqr[X_\eps]{(H_\eps-\im)u}
  \end{equation*}
  for $u \in \dom H_\eps \subset \dom \qf h_\eps = \Sob {X_\eps}$ and
  $0 < \eps \le \eps_0$.
\end{lemma}

%----------------------------------------------------------------------
%
\section{Approximation of $\delta$-couplings}
\label{sec:delta}
%
%----------------------------------------------------------------------

After this preliminaries we can pass to our main problems.  The first
one concerns approximation of a $\delta$-coupling by Schr\"odinger
operators with scaled potentials supported by the vertex regions.  For
the sake of simplicity most part of the discussion will be done for
the situation with a single vertex as described in \Sec{graph}.

%----------------------------------------------------------------------
\subsection{Quasi-unitary operators}
\label{sec:quasi.unitary}
%----------------------------------------------------------------------
First we define quasi-unitary operators mapping from $\HS$ to $\wt
\HS$ and vice versa, as well as their analogues on the scales of
order one, namely $\HS^1$ and $\wt \HS^1$.  Here,
% ------------- %
\begin{align}
  \label{eq:spaces}
    \HS       &:= \Lsqr G, &
    \HS^1     &:= \Sob  G, &
    \wt \HS     &:= \Lsqr {X_\eps}, &
    \wt \HS^1   &:= \Sob  {X_\eps}.
\end{align}
 % ------------- %
Moreover, we need a relation between the different constants of
the graph and the manifold model introduced above.  Specifically,
we set
\begin{equation}
  \label{eq:def.p}
  p_e:= (\vol_{d-1} Y_e)^{1/2} \Und
  q(v) = \int_{X_v} Q \dd x_v.
\end{equation}
Let $\map J \HS {\wt \HS}$ be given by
 % ------------- %
\begin{equation}
  \label{eq:def.j}
  J f := \eps^{-{(d-1)/2}}\bigoplus_{e \in E} (f_e \otimes \dashone_e)
           \oplus 0
\end{equation}
 % ------------- %
with respect to the decomposition~\eqref{eq:lsqr.xeps}.  Here
$\dashone_e$ is the normalised eigenfunction of $Y_e$ associated
to the lowest (zero) eigenvalue, i.e.
$\dashone_e(y)=(\vol_{d-1}Y_e)^{-1/2}$.   In order to relate the
Sobolev spaces of order one we need a similar map: we define $\map
{J^1} {\HS^1} {\wt \HS^1}$ by
 % ------------- %
\begin{equation}
  \label{eq:j.1}
  J^1 f := \eps^{-(d-1)/2} \Bigl(\bigoplus_{e \in E} (f_e \otimes \dashone_e)
           \oplus f(v) \1_v \Bigr),
\end{equation}
 % ------------- %
where $\1_v$ is the constant function on $X_v$ with value $1$.   Note
that the latter operator is well defined:
\begin{equation*}
  (J^1f)_e(0,y)=\eps^{-(d-1)/2} p_e^{-1} f_e(0)
  = \eps^{-(d-1)/2} f(v)
  = (J^1f)_v(x)
\end{equation*}
for any $x \in X_v$ due to~\eqref{eq:def.p} and~\eqref{eq:eval.f},
i.e., the function $J^1f$ matches along the different components
of the manifold,  thus $Jf \in \Sob{X_\eps}$.  Moreover, $f(v)$ is
defined for $f \in \Sob G$ (see \Lem{sob.tr}).

The mapping in the opposite direction, $\map {J'} {\wt \HS} \HS$,
is given by the adjoint, $J':= J^*$, which means that
 % ------------- %
\begin{equation}
  \label{eq:j.}
  (J' u)_e (s) = \eps^{(d-1)/2} \iprod[Y_e]{\dashone_e}{u_e(s,\cdot)}
  = \eps^{(d-1)/2} p_e \avint_e u(s).
\end{equation}
 % ------------- %
Furthermore, we define $\map {J'{}^1} {\wt \HS^1}
{\HS^1}$ by
 % ------------- %
\begin{equation}
  \label{eq:j.1.}
  (J'_e{}^{1} u)(s):=
  \eps^{(d-1)/2} \Bigl[ \iprod[Y_e]{\dashone_e}{u_e(s,\cdot)}
       + \chi_e(s) p_e \Bigl(\avint_v u - \avint_e u(0) \Bigr)
             \Bigr].
\end{equation}
 % ------------- %
Here $\chi_e$ is a smooth cut-off function such that $\chi_e(0)=1$
and $\chi_e(\ell_e)=0$.   If we choose the function $\chi_e$ to be
piecewise affine linear with $\chi_e(0)=1$, $\chi_e(a)=0$ and
$\chi_e(\ell_e)=0$, then $\normsqr[I_e]{\chi_e}=a/3\le a$ and
$\normsqr[I_e]{\chi_e'}=a^{-1}$.   Moreover, $(J'_e{}^1 u)_e(0)=
\eps^{(d-1)/2} p_e \avint_v u$ so that $f:= J'_e{}^{1} u$
satisfies $\ul f(0) \in \C \ul p$, and therefore $f \in \Sobx {\ul
p} G$.  Note that by construction of the manifold, we have
$\avint_{\bd e X_v}u = \avint_e u(0)$.

%----------------------------------------------------------------------
\subsection{Closeness assumptions}
\label{sec:closeness}
%----------------------------------------------------------------------

Let us start this subsection with a lower bound on the operators $H$
and $H_\eps$ in terms of the model parameters; for the definitions of
the constants $C_{1/2}$, $\eps_{1/2}$ and $\wt C_{1/2}$ see
\Lem{delta.pert} and \Lem{ham.pert}.  Note that $\wt C_{1/2}$ still
depends on $\norm[\infty] Q$ and $\ell_-$.
%----------------------------------------------------------------------
\begin{lemma}
  \label{lem:lower.bd}
  For $\eps \in (0, \eps_{1/2}]$ the operators $H_\eps$ and $H$ are
  bounded from below by $\lambda_0:=-\wt C_{1/2}$.  Moreover, if all
  lengths are finite, i.e.\ $\ell_e<\infty$, and $q(v) \le 0$, then we
  have
  \begin{equation*}
    \inf \spec{H} \le \frac {q(v)}{\vol X_E} \Und
    \inf \spec{H_\eps}
    \le%  \frac {\int_{X_v} Q \dd x_v}
%               {\vol X_E + \eps \vol X_v}
%     =
    \frac {q(v)}
              {\vol X_E + \eps \vol X_v},
  \end{equation*}
  where $X_E := \bigdcup_e X_e$ is the union of the edge
  neighbourhoods.
\end{lemma}
%----------------------------------------------------------------------
\begin{proof}
  We have to calculate the maximum of $C_{1/2}$ and $\wt C_{1/2}$. Due
  to~\eqref{eq:def.p} we have $\abssqr{\ul p} = \vol \bd X_v$ and
  $\abs {q(v)} = \bigabs{\int_{X_v} Q \dd x_v} \le \norm[\infty] Q
  \vol X_v$ so that
  \begin{equation}
    \label{eq:c.1-2}
    C_{1/2}
    \le \max \Bigl\{
          4 \cvol^2 \normsqr[\infty] Q,
          \frac{2\cvol \norm[\infty] Q}{\ell_-}
       \Bigr\}
    \le \wt C_{1/2}
     = \max \Bigl\{
            64 \cvol^2 \normsqr[\infty] Q,
                 \frac {8\cvol \norm[\infty] Q}{\ell_-}
            \Bigr\},
  \end{equation}
  where $\cvol:=\vol X_v/\vol\bd X_v$.  The spectral estimates then follow
  by inserting suitable test functions into the Rayleigh quotients $\qf
  h(f)/\normsqr f$ and $\qf h_\eps(u)/\normsqr u$.  For $f$, we choose
  the edgewise constant function $f_e(x)=p_e$.  Note that $f \in \Sobx
  {\ul p} G$. On the manifold, we choose the constant $u:=J^1f =
  \eps^{(d-1)/2} \1$. The upper bound on the infimum on the spectrum
  follows by the relation $\ell_e p_e^2 = \vol X_e$
  using~\eqref{eq:def.p}.
\end{proof}
%----------------------------------------------------------------------

Now we are in position to demonstrate that the two Hamiltonians are
close to each other. We start with estimates of the identification
operators and the forms $\qf h$, $\qf h_\eps$ in terms of the ``free''
quadratic forms $\qf d$ and $\qf d_\eps$:
%----------------------------------------------------------------------
\begin{lemma}
  \label{lem:closeness}
  The identification operators $J$, $J'=J^*$, $J^1$, $J'{}^1$ and the
  quadratic forms $\qf h_\eps$ and $\qf h$ fulfil the estimates
  \begin{subequations}
    \label{eq:closeness}
    \begin{align}
      \label{eq:j1}
      \normsqr{Jf - J^1f} &\le \delta_\eps^2 \normsqr[\Sob G] f, &
      \normsqr{J'u - J'{}^1u} &\le \delta_\eps^2 \normsqr[\Sob {X_\eps}] u,\\
      \label{eq:j.bdd}
      \normsqr{J f} &= \normsqr f, &
      \normsqr{J' u} &\le \normsqr u,\\
      \label{eq:j.inv}
      J' J f &= f,&
      \normsqr{J J' u - u} &\le \delta_\eps^2 \normsqr[\Sob{X_\eps}] u,\\
      \label{eq:j.comm.1}
      \bigabs{ \qf h(J'{}^1 u, f) - \qf h_\eps(u, J^1 f)}
       &\le \delta_\eps \norm[\Sob {X_\eps}] u \norm[\Sob G] f
    \end{align}
  \end{subequations}
  with $\delta_\eps=\Err(\eps^{1/2})$ as $\eps\to 0$, being given
  explicitly by
  \begin{multline}
  \label{eq:def.delta}
    \delta_\eps^2 := \max \Bigl\{ \frac{8 \eps \cvol} {\ell_0},
         \frac {\eps^2} {\lambda_2(E)},
     4 \eps^2 \Bigl[ \frac 1 {\lambda_2(v)}
       + \cvol \Bigl( 1 + \frac 2 {\ell_0 \lambda_2(v)}\Bigr)\Bigr],\\
      \frac{2 \eps}{\ell_0} \Bigl(1 + \frac 2 {\ell_0 \lambda_2(v)} \Bigr),
    \frac {4 \eps \cvol \normsqr[\infty] Q} {\ell_0 \lambda_2(v)}\Bigr\}.
  \end{multline}
  Here, $\ell_0 := \min \{1,\ell_-\}=\min_e \{1,\ell_e\} \le 1$,
  $\lambda_2(E) := \min_e \lambda_2(e)$ and $\cvol = \vol X_v/\vol\bd
  X_v$. Moreover, $\lambda_2(e)$ and $\lambda_2(v)$ denote the second
  (first non-vanishing) eigenvalue of the (Neumann-)Laplacian on $Y_e$
  and $X_v$, respectively.
\end{lemma}
%----------------------------------------------------------------------
\begin{proof}
  The first condition in~\eqref{eq:j1} is here
  % ------------- %
  \begin{equation*}
    \normsqr[X_\eps]{Jf - J^1f}
    = \eps \vol X_v \abssqr{f(v)} \le
    \eps \cvol
    \Bigl( \normsqr[G]{f'}
    + \frac 2 {\ell_0} \normsqr[G] f
    \Bigr)
  \end{equation*}
 % ------------- %
  using \Lem{sob.tr} with $a=\ell_0$ and the fact that $\abssqr {\ul
    p} = \vol \bd X_v$ due to~\eqref{eq:def.p}.  Next we need to show
  the second estimate in~\eqref{eq:j1}.  In our situation, we have
  % ------------- %
  \begin{equation*}
    \normsqr[G]{J' u - J'{}^1 u}
    = \eps^{d-1} \sum_{e \in E}
           \normsqr[I_e]{\chi_e} p_e^2
           \bigabssqr{\avint_v u - \avint_e u(0)}
    \le \eps \Bigl(1 + \frac 2 {\ell_0 \lambda_2(v)}
                   \Bigr) \normsqr[X_\vxeps]{\de u}
  \end{equation*}
  % ------------- %
  using \Lem{av.int} with $a=\wt a=\ell_0$.
  Moreover,~\eqref{eq:j.bdd} and the first equation
  in~\eqref{eq:j.inv} are easily seen to be fulfilled. The second
  estimate in~\eqref{eq:j.inv} is more involved.  Here, we have
  % ------------- %
  \begin{equation*}
    \normsqr{J J' u - u} =
    \sum_e  \normsqr[X_\edeps] {u - \avint_e u}
    + \normsqr[X_\vxeps] u.
  \end{equation*}
  % ------------- %
  The first term can be estimated as in~\eqref{eq:min-max} by
  \begin{equation*}
    \bignormsqr[X_\edeps] {u - \avint_e u}
    = \int_{I_e} \bignormsqr[Y_e] {u(s) - \avint_e u(s)} \dd s
    \le \frac 1 {\lambda_2(e)}
          \int_{I_e} \normsqr[Y_e] {\de_{Y_e}u(s)} \dd s
   = \frac {\eps^2} {\lambda_2(e)}
          \normsqr[X_\edeps] {\de_{Y_e}u},
  \end{equation*}
  where $u(s):= u(s,\cdot)$. The second term can be estimated by
  \Lem{vx.est}, so that
  \begin{equation*}
    \delta_\eps^2 \ge \max
    \Bigl\{
         4 \eps^2 \Bigl[ \frac 1 {\lambda_2(v)}
         + \cvol \Bigl( 1 + \frac 2 {\ell_0 \lambda_2(v)}\Bigr)\Bigr],
         \frac {\eps^2} {\lambda_2(E)},
%         4\eps \cvol,   % contained estimated by next constant ...
         \frac{8 \eps \cvol} {\ell_0}
    \Bigr\},
  \end{equation*}
  which is sufficient for the estimate~\eqref{eq:j.inv}.

  Let us finally prove~\eqref{eq:j.comm.1} in our model. Note that
  this estimate differs from the ones given in~\cite{post:06} by the
  absence of the potential term $Q_\eps=\eps^{-1}Q$ there. In our
  situation, we have
  % ------------- %
  \begin{multline*}
    \bigabssqr{\qf h(J'{}^1 u, f) - \qf h_\eps(u, J^1 f)}\\
    \le 2 \eps^{d-1} \Bigl[ \Bigabssqr{\sum_e p_e
       \Bigl(\avint_v \conj u - \avint_e \conj u(0)
       \Bigr)
       \iprod[I_e]{\chi_e'} {f'}}
    + \bigabssqr{
           q(v) \avint_v \conj u - \iprod[X_v]{Qu} {\1_v}}
       \abssqr {f(v)}
    \Bigr].
  \end{multline*}
  % ------------- %
  Note that the derivative terms cancel on the edges due to the
  product structure of the metric and the fact that $\de_{Y_e}
  \dashone_e =0$ and the vertex contribution vanishes due to
  $\de_{X_v} \1 = 0$. The first term can be estimated by
  \begin{equation*}
    2 \eps \Bigl(\wt a + \frac 2 {\wt a \lambda_2(v)} \Bigr)
   \frac 1 a \normsqr[X_\vxeps] {\de u}
   \le \frac {2\eps}{\ell_0}
      \Bigl(1 + \frac 2 {\ell_0 \lambda_2(v)} \Bigr)
  \end{equation*}
  using Cauchy-Schwarz, \Lem{av.int} and the fact that
  $\normsqr[I_e]{\chi_e'}=1/a \le 1/\ell_0$ by our choice of $\chi_e$.
  For the second term, we use our definition $q(v)=\int_{X_v} Q \dd
  x_v$ and $q(v) \avint_v \conj u = \iprod[X_v] u {\avint_v Q \1_v}$
  to conclude
  % ------------- %
  \begin{multline*}
    \bigabssqr{q(v) \avint_v \conj u - \iprod[X_v]{Qu} {\1_v}}
    = \bigabssqr{\bigiprod[X_v] u {\avint_v Q - Q}}\\
    = \bigabssqr{\iprod[X_v] u {P_v Q}}
    = \bigabssqr{\iprod[X_v] {P_v u} Q}
    \le \frac 1 {\lambda_2(v)} \normsqr[X_v]{\de u} \normsqr[X_v] Q
  \end{multline*}
  % ------------- %
  where $P_v u := u - \avint_v u$ is the projection onto the
  orthogonal complement of $\1_v$. The last estimate follows
  from~\eqref{eq:min-max}. Collecting the error terms for the
  sesquilinear form estimate, we obtain
  \begin{equation*}
    \delta_\eps^2 \ge \max
    \Bigl\{
        \frac{2 \eps}{\ell_0}
            \Bigl(1 + \frac 2 {\ell_0 \lambda_2(v)} \Bigr),
        \frac {4\eps \cvol \normsqr[\infty] Q} {\ell_0 \lambda_2(v)}
    \Bigr\}
  \end{equation*}
  as lower bound on $\delta_\eps$, using also \Lem{sob.tr} for the
  estimate on $\abssqr{f(v)}$, and $\normsqr[X_v] Q \le \vol X_v
  \normsqr[\infty] Q$.
\end{proof}
%----------------------------------------------------------------------

Now we can prove our main result on the approximation of a
$\delta$-coupling in the manifold model; for more details on the
notion of ``$\delta$-closeness'' we refer to~\cite[App.]{post:06}.
The resolvent estimate at $z=\im$ will be needed in \Sec{delta'} when
the lower bound $\lambda_0$ depends on $\eps$ and may tend to
$-\infty$ as $\eps \to 0$.  Recall the definition of $\wt C_{1/2}$,
$0<\eps_{1/2}$ (see~\eqref{eq:def.eps.eta}) and $\lambda_0:= -\wt
C_{1/2}$, and that $\wt C_{1/2} \ge C_{1/2}$.
%----------------------------------------------------------------------
\begin{theorem}
  \label{thm:closeness}
  For $\eps \in (0, \eps_{1/2}]$, the operators $H_\eps-\lambda_0$ and
  $H-\lambda_0$ are $\sqrt 2\delta_\eps$-close with
  $\delta_\eps=\Err(\eps^{1/2})$ given in~\eqref{eq:def.delta}; in
  other words, there is an identification operator $\map J {\Lsqr
    G}{\Lsqr{X_\eps}}$ such that $J^*J=\id$,
  \begin{equation*}
    \bignorm{(\id - J J^*)(H_\eps-\lambda_0)^{-1}}
    \le \sqrt 2 \delta_\eps
      \und
    \bignorm{J(H-\lambda_0)^{-1} - (H_\eps - \lambda_0)^{-1}J}
    \le 3\sqrt 2 \delta_\eps.
  \end{equation*}
  Moreover, for $\eps \in (0, \eps_{1/2}]$ we have the estimate
  \begin{equation*}
    \bignorm{J(H-\im)^{-1} - (H_\eps - \im)^{-1}J}
    \le 10 \delta_\eps \max \{\wt C_{1/2}, \sqrt 2\},
  \end{equation*}
  where $\norm \cdot$ denotes the operator norm for operators from
  $\Lsqr G$ into $\Lsqr{X_\eps}$.
\end{theorem}
%----------------------------------------------------------------------
\begin{proof}
  The closeness of the operators $H - \lambda_0$ and $H_\eps -
  \lambda_0$ follows from the estimate
  \begin{equation*}
%    \normsqr[X_\eps] {(\lapl_{X_\eps} + 1)^{1/2}u}
    %=
    \normsqr[X_\eps] {\de u} + \normsqr[X_\eps ]u%\\
    \le 2 \bigl( \qf h_\eps (u)
               + (1- \lambda_0) \normsqr[X_\eps] u \bigr)
     = 2 \normsqr[X_\eps]{(H_\eps - \lambda_0 + 1)^{1/2} u}
  \end{equation*}
  by \Cor{ham.pert}, and similarly for $H$ on $G$ by \Cor{delta.pert}
  and~\eqref{eq:c.1-2}, together with \Lem{closeness}.  The resolvent
  estimate can be seen as follows: Let $R:=(H-\im)^{-1}$ and
  $R:=(H_\eps-\im)^{-1}$, and let $\wt f \in \Lsqr G$, $\wt u \in
  \Lsqr {X_\eps}$. Setting $f := R \wt f \in \dom H$ and $u := R_\eps
  \wt u \in \dom H_\eps$, we have
  \begin{multline*}
    \iprod{\wt u} {(JR-R_\eps J)\wt f}
    = \iprod {\wt u} {Jf} - \iprod u {J \wt f}\\
    = \iprod {\wt u} {(J-J^1)f}
     +  \bigl( \qf h_\eps(u,J^1 f)
               -\qf h(J'{}^1 u, f) \bigr)
     +\iprod {(J'{}^1- J^*) u} {\wt f}\\
     - \im \bigl( \iprod u {(J^1 - J)f}
                 +\iprod {(J'{}^1- J^*) u} f\bigr),
  \end{multline*}
  and therefore
  \begin{equation*}
    \bigabs{\iprod{\wt u} {(JR-R_\eps J)\wt f}}
    \le 10 \delta_\eps \max \{\wt C_{1/2}, \sqrt 2\}
                  \norm {\wt f} \norm {\wt u}
  \end{equation*}
  using \Lems{res.est}{res.est.mfd}, and the fact that $C_{1/2} \le \wt
  C_{1/2}$.
\end{proof}
%----------------------------------------------------------------------

Using the abstract results of~\cite[App.~A]{post:06}
or~\cite{post:pre08a}, we can show the resolvent convergence and the
convergence other functions of the operator:
\begin{theorem}
  \label{thm:res}
  We have
  \begin{subequations}
  \label{eq:main.res}
    \begin{gather}
      \norm{J(H - z)^{-1} - (H_\eps - z)^{-1}J}
      = \Err(\eps^{1/2}),\\
      \norm{J(H - z)^{-1}J' - (H_\eps - z)^{-1}} = \Err(\eps^{1/2})
    \end{gather}
  \end{subequations}
  for $z \notin [\lambda_0,\infty)$. The error depends only on
  $\delta_\eps$, given in~\eqref{eq:def.delta}, and on $z$.  Moreover,
  we can replace the function $\phi(\lambda)=(\lambda-z)^{-1}$ by any
  measurable, bounded function converging to a constant as $\lambda
  \to \infty$ and being continuous in a neighbourhood of $\spec H$.
\end{theorem}

The following spectral convergence is also a consequence of the
$\Err(\eps^{1/2})$-closeness; for details of the uniform
convergence of sets, i.e.\ the convergence in Hausdorff-distance
sense we refer to~\cite[App.~A]{herbst-nakamura:99}
or~\cite{post:pre08a}.
 % ------------- %
\begin{theorem}
  \label{thm:spec}
  The spectrum of $H_\eps$ converges to the spectrum of $H$ uniformly
  on any finite energy interval. The same is true for the essential
  spectrum.
\end{theorem}
 % ------------- %
\begin{proof}
  The spectral convergence is a direct consequence of the closeness,
  as it follows from the general theory developed
  in~\cite[Appendix]{post:06} and~\cite{post:pre08a}.
\end{proof}
 % ------------- %
 For the discrete spectrum we have the following result:
 % ------------- %
\begin{theorem}
  \label{thm:disc.spec}
  For any $\lambda \in \disspec H$ there exists a family
  $\{\lambda_\eps\}_\eps$ with $\lambda_\eps \in \disspec {H_\eps}$
  such that $\lambda_\eps \to \lambda$ as $\eps \to 0$. Moreover, the
  multiplicity is preserved. If $\lambda$ is a simple eigenvalue with
  normalised eigenfunction $\phi$, then there exists a family of
  simple normalised eigenfunctions $\{\phi_\eps\}_\eps$ of $H_\eps$
  ($\eps$ small) such that
 % ------------- %
  \begin{equation*}
    \norm[X_\eps]{J\phi - \phi_\eps} \to 0
  \end{equation*}
 % ------------- %
  as $\eps \to 0$.
\end{theorem}
 % ------------- %
We remark that the convergence of higher-dimensional eigenspaces
is also valid, however, it requires some technicalities which we
skip here.

To summarise, we have shown that the $\delta$-coupling with
weighted entries can be approximated by a geometric setting and a
potential located on the vertex neighbourhood.

Let us briefly sketch how to extend the above convergence results
\ThmS{closeness}{disc.spec} to more complicated --- even to
non-compact --- graphs. Denote by $G$ a metric graph, given by the
underlying discrete graph $(V,E,\bd)$ with $\map \bd E {V \times V}$,
$\bd e = (\bd_-e,\bd_+e)$ denoting the initial and terminal vertex,
and the length function $\map \ell E {(0,\infty)}$, such that each
edge $e$ is identified with the interval $I_e=(0,\ell_e)$ (for
simplicity, we assume here that all length are finite, i.e., $\ell_e <
\infty$). Let $X_\eps$ be the corresponding approximating manifold
constructed from the building blocks $X_\edeps=I_e \times \eps Y_e$
and $X_\vxeps=\eps X_v$ as in \Sec{mfd}. For more details, we refer
to~\cite{exner-post:05,post:06,exner-post:08a,post:pre08a}. Since a
metric graph can be constructed from a number of star graphs with
identified end points of the free ends, we can define global
identification operators. We only have to assure that the global error
we make is still uniformly bounded:
\begin{theorem}
  \label{thm:non-compact}
  Assume that $G$ is a metric graph and $X_\eps$ the corresponding
  approximating manifold constructed according to $G$. If
  \begin{equation*}
    \inf_{v \in V} \lambda_2(v) > 0, \quad
    \sup_{v \in V} \frac{\vol X_v}{\vol \bd X_v} < \infty, \quad
    \sup_{v \in V} \norm[\infty] {Q \restr{X_v}} < \infty, \quad
    \inf_{e \in E} \lambda_2(e) > 0, \quad
    \inf_{e \in E} \ell_e > 0,
  \end{equation*}
  then the corresponding Hamiltonians $H=\laplacian G + \sum_v q(v)
  \delta_v$ and $H_\eps = \laplacian{X_\eps} + \sum_v \eps^{-1} Q_v$
  are $\delta_\eps$-close, where the error
  $\delta_\eps=\Err(\eps^{1/2})$ depends only on the above indicated global
  constants.
\end{theorem}
%----------------------------------------------------------------------
%
\section{Approximation of the $\delta'_\mathrm{s}$-couplings}
\label{sec:delta'}
%
%----------------------------------------------------------------------

The main aim of this section is to show how a the symmetrised
$\delta'$-coupling, or $\delta'_\mathrm{s}$, can be approximated
using manifold model discussed above. To this aim we shall use a
result of~\cite{cheon-exner:04} by which a
$\delta'_\mathrm{s}$-coupling can be approximated by means of
several $\delta$-couplings on the same metric graph, located close
to the vertex and ``lift'' this approximation to the manifold. For
the sake simplicity we will again consider the star-shape setting
with a single vertex. We want to stress, however, that the method
we use can be directly generalised to more complicated graphs but
also, what is equally important, to other vertex couplings, once
they can be approximated by combinations of $\delta$-couplings on
the graph, possibly with an addition of extra edges ---
see~\cite{exner-turek:05,exner-turek:07}.

Let thus $G=I_{v_0}$ be a star graph as in \Sec{model} where we
denote the vertex in the centre by $v_0$ and where we label the
$n=\deg v$ edges by $e=1,\dots, n$. Again for simplicity, we
assume that all the (unscaled) transversal volumes $p_e^2=\vol
Y_e$ are the same; without loss of generality we may put $\vol
Y_e=1$. Moreover, we assume that all lengths are finite, i.e.\
$\ell_e < \infty$, and equal, so we may put $\ell_e=1$. First we
recall the definition of the $\delta'_{\mathrm s}$-coupling: the
operator $H^\beta$, formally written as $H^\beta = \lapl_{G} +
\beta \delta_{v_0}'$, acts as $(H^\beta f)_e = -f''_e$ on each
edge for functions $f$ in the domain
\begin{multline}
  \label{eq:h.beta}
  \dom H^\beta :=
    \Bigset {f \in \Sobx[2] \max G}
      {\forall e_1,e_2 \colon f'_{e_1}(0)=f'_{e_2}(0)=:f'(0),  \;
       \sum_e f_e(0)=\beta f'(0),\\
     \forall e \colon f'_e(\ell_e)=0}.
\end{multline}
For the sake of definiteness we imposed here Neumann conditions at
the free ends of the edges, however, the choice is not
substantial; we could use equally well Dirichlet or any other
boundary condition. The corresponding quadratic form is given as
\begin{equation*}
  \qf h^\beta(f) =\sum_e \normsqr{f_e'}
  + \frac 1 \beta \Bigabssqr{\sum_e f_e(0)},
  \qquad
  \dom \qf h^\beta = \Sobx \max G
\end{equation*}
if $\beta \ne 0$ and
\begin{equation*}
  \qf h^\beta(f) =\sum_e \normsqr{f_e'},
  \qquad
  \dom \qf h^\beta =
  \bigset{f \in \Sobx \max G} {\sum_e f_e(0)=0}
\end{equation*}
if $\beta=0$; the condition $f \in H^0$ is obviously dual to the free
(or Kirchhoff) vertex coupling --- see,
e.g.,~\cite[Sec.~3.2.3]{kuchment:04}.

 The (negative) spectrum of
$H^\beta$ is easily found:
\begin{lemma}
  \label{lem:spec.delta'}
  If $\beta \ge 0$ then $H^\beta \ge 0$. On the other hand, if $\beta<0$ then
  $H^\beta$ has exactly one negative eigenvalue $\lambda=-\kappa^2$
  where $\kappa$ is the solution of the equation
  \begin{equation}
    \label{eq:kappa.beta}
    \cosh \kappa + \frac {\beta \kappa}{\deg v} \sinh \kappa = 0.
  \end{equation}
\end{lemma}
\begin{proof}
  The non-negativity of $H^\beta$ follows from the quadratic form
  expression for $\beta>0$ and $\beta=0$.  We make the ansatz
  \begin{equation*}
    f_e(s)=\cosh \kappa(1-s)
  \end{equation*}
  fulfilling automatically the Neumann condition at $s=1$ and the
  continuity condition at $s=0$ since $f_e'(0)=-\kappa \sinh \kappa$
  is independent of $e$. The remaining condition at zero leads to the
  above relation of $\kappa$ and $\beta$, showing in another way that if
  $\beta \ge 0$ there cannot exist a negative eigenvalue.
\end{proof}

The main idea of the approximation of a
$\delta'_\mathrm{s}$-coupling by Schr\"odinger operators on a
manifold is to employ a combination of $\delta$-couplings in an
operator one may call an \emph{intermediate Hamiltonian}
$H^{\beta,a}$, and then to use the approximations for
$\delta$-couplings given in the previous section.

\begin{figure}[h]
  \centering
%----------------------------------------------------------------------
%  \input{pot-approx-fig1.pstex_t}
%----------------------------------------------------------------------
\begin{picture}(0,0)%
\includegraphics{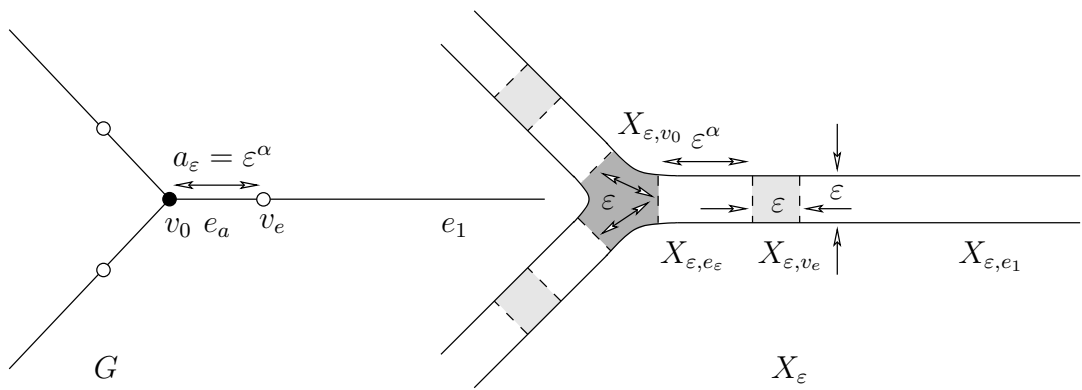}%
\end{picture}%
\setlength{\unitlength}{4144sp}%
\begin{picture}(6431,2281)(79,-1565)
\put(4650,-1496){$X_\eps$}%
\put(598,-1496){$G$}%
\put(1070,-227){$a_\eps=\eps^\alpha$}%
\put(1021,-622){$v_0$}%
\put(1586,-594){$v_e$}%
\put(1256,-622){$e_a$}%
\put(2676,-622){$e_1$}%
\put(5000,-424){$\eps$}%
\put(4171,-114){$\eps^\alpha$}%
\put(3635,-481){$\eps$}%
\put(4650,-500){$\eps$}%
\put(4569,-807){$X_{\eps,v_e}$}%
\put(3986,-807){$X_{\eps,e_\eps}$}%
\put(5770,-807){$X_{\eps,e_1}$}%
\put(3731,-033){$X_{\eps,v_0}$}%
\end{picture}%
%----------------------------------------------------------------------
  \caption{The intermediate graph picture used in the
    $\delta'_\mathrm{s}$-approximation and the corresponding manifold
    model.}
\label{fig:pot-approx-fig1}
\end{figure}

In order to define $H^{\beta,a}$, we first modify the (discrete)
structure of the graph $G$ inserting additional vertices $v_e$ of
degree $2$ on the edge $e$ with the distance $a \in (0,1)$ from the
central vertex $v_0$ (see \Fig{pot-approx-fig1}).  Each edge $e$ is
splitted into two edges $e_a$ and $e_1$.  We denote the metric graph
with the additional vertices $v_e$ and splitted edges by $G_a$, i.e.,
$V(G_a)=\{v_0\} \cup \set{v_e}{e = 1,\dots, n}$,
$E(G_a)=\set{e_a,e_1}{e =1,\dots n}$ and $\ell_{e_a}=a$,
$\ell_{e_1}=1-a$. This metrically equivalent graph $G_a$ will be
needed when associating the corresponding manifold.
\begin{remark}
  It is useful to note that the Laplacians $\laplacian G$ and
  $\laplacian {G_a}$ associated to the metric graphs $G$ and $G_a$ are
  unitarily equivalent. Indeed, introducing additional vertices of degree
  two does not change the original quadratic form $\qf d_G$ with the domain
  $\Sob G=\dom \qf d$ associated to the free operator $\laplacian G =
  H_{(G,0)}$. Figuratively speaking, the free operator does not see these
  vertices of degree two. We
  just have to change the coordinate on the edge $e$, i.e.\ we can
  either use the original coordinate $s \in (0,\ell_e)$ on the edge
  $e$ or we can split the edge $e$ into two edges $e_a$ and $e_1$ of
  length $\ell_{e_a}=a$ and $\ell_{e_1}=\ell_e-a=1-a$ with the
  corresponding coordinates.
\end{remark}
The core of the approximation lies in a suitable, $a$-dependent
choice of the parameters of these $\delta$-couplings. Writing the
operator in terms of the formal notation introduced
in~\eqref{eq:delta}, we put
\begin{equation*}
   H^{\beta,a} := \lapl_G + b(a) \delta_{v_0} + \sum_e c(a) \delta_{v_e},
  \qquad
  b(a) = - \frac \beta{a^2}, \qquad
  c(a) = - \frac 1 a,
\end{equation*}
to be the \emph{intermediate} Hamiltonian. Notice that the
strength of central $\delta$-coupling depends on $\beta$ while the
added $\delta$-interactions are attractive, the sole parameter
being the distance $a$. The operator can be defined via its
quadratic form
\begin{equation*}
   {\qf h}^{\beta,a}(f) := \sum_e \normsqr{f_e'}
     - \frac \beta{a^2} f(0) - \frac 1 a \sum_e\abssqr{f_e(a)},
  \qquad
  \dom  {\qf h}^a = \Sob G,
\end{equation*}
where $\Sob G=\Sobx {\ul p} G$ with $\ul p=(1,\dots,1)$, i.e.\ the
functions $f\in \Sob G$ are distinguished by being continuous at
$v_0$, $\:f_{e_1}(0)=f_{e_2}(0)=:f(0)$.

The next theorem shows that the intermediate Hamiltonian converges
indeed to the $\delta'_\mathrm{s}$-coupling with the strength
$\beta$ on the star-shaped graph:
\begin{theorem}[Cheon, Exner]
  \label{thm:delta'}
  We have
  \begin{equation*}
    \norm{( H^{\beta,a} - z)^{-1} - (H^\beta - z)^{-1}} =\Err(a)
  \end{equation*}
  as $a \to 0$ for $z \notin \R$, where $\norm \cdot$ denotes the
  operator norm on $\Lsqr G$.\footnote{The claim made
    in~\cite{cheon-exner:04} is only that the norm tends to zero,
    however, the rate with which it vanishes is obvious from the
    proof. We remove the superfluous $\deg v$ from the definition of
    $H^{\beta,a}$ in that paper. It should also be noted that the
    proof in~\cite{cheon-exner:04} is given for star graphs with
    semi-infinite edges but the argument again modifies easily to the
    finite-length situation we consider for convenience here.}
\end{theorem}
Note that the choice of the parameters $b(a)$ $c(a)$ of the
$\delta$-interactions as functions of the distance $a$ follows from a
careful analysis of the resolvents of $H^{\beta,a}$ and $H^\beta$.
Each of these is highly singular as $a\to 0$, however, in the
difference all the singularities cancel leaving us with a vanishing
expression. Needless to say, that such a limiting process is highly
non-generic.

Let us now consider the manifold model approaching the intermediate
situation Hamiltonian $ H^{\beta,a}$ in the limit $\eps\to 0$ with
$a=a_\eps=\eps^\alpha$ and $0<\alpha <1$ to be specified later on. Let
$X_\eps$ be a manifold model of the graph $G$ as shown in
\Fig{pot-approx-fig1}. For the additional vertices of degree two we
choose the vertex neighbourhoods as a part of the cylinder of length
$\eps$ and distance of order of $a_\eps$ from the central vertex
$v_0$. The edge $e_{a_\eps}=:e_\eps$ now has the length
$a_\eps=\eps^\alpha$ depending on $\eps$.  The ``free'' edge $e_1$
joining $v_e$ with the free end point at $s=1$ is again
$\eps$-depending, namely it has the length $1-a_\eps=1-\eps^\alpha$.
By the argument given in \Lem{long.err} we can deal with this error
and assume that this edge again has length one, the price being an
extra error of order $\Err(\eps^\alpha)$, affecting neither the final
result nor the quantitative error estimate. Next we have to choose the
potentials in the vicinity of the vertices $v=v_0$ and $v=v_e$. The
simplest option is to assume that they are constant,
\begin{equation*}
  Q_\vxeps (x) := \frac 1 \eps \cdot
           \frac {q_\eps (v)}{\vol X_v}, \qquad x \in X_v
\end{equation*}
so that $\int_{X_v} Q_\vxeps \dd x = \eps^{-1}q_\eps(v)$
(see~\eqref{eq:pot.eps} and~\eqref{eq:def.p}), where we put
\begin{equation*}
  q_\eps(v_0) := b(\eps^\alpha)
    = - \beta \eps^{-2\alpha}
    \Und
  q_\eps(v_e) := c(\eps^\alpha)
    = - \eps^{-\alpha}.%\frac \beta {\deg v} \eps^{-2\alpha}
\end{equation*}
The corresponding manifold Hamiltonian and the respective
quadratic form are then given by
\begin{equation}
  \label{eq:h.beta.eps}
  H_\eps^\beta
  = \laplacian{X_\eps}
  - \eps ^{-1-2\alpha} \frac \beta {\vol X_{v_0}} \1_{X_{v_0}}
  - \eps^{-1-\alpha} \sum_{e \in E} \1_{X_{v_e}}
\end{equation}
and
\begin{equation*}
  \qf h_\eps^\beta(u)
  = \normsqr[X_\eps]{\de u}
  - \eps ^{-1-2\alpha} \frac \beta {\vol X_{v_0}} \normsqr[X_{\eps,v_0}] u
  - \eps^{-1-\alpha} \sum_{e \in E} \normsqr[X_{\eps,v_e}] u,
\end{equation*}
respectively. Note that the unscaled vertex neighbourhood $X_{v_e}$ of
the added vertex $v_e$ has volume $1$ by construction.

Before proceeding to the approximation itself, let us first make
some comments about the lower bounds of the operators $
H^{\beta,a}$ and their manifold approximations $H_\eps^\beta$:
\begin{lemma}
  \label{lem:lower.bd.gr}
  If $\beta < 0$, then the spectrum of $ H^{\beta,a}$ is uniformly
  bounded from below as $a \to 0$, in other words, there is a constant $C>0$
  such that
  \begin{equation*}
    \inf  \spec{ H^{\beta,a}}
    \ge
    -C \quad\text{as} \quad a \to 0.
  \end{equation*}
  If $\beta \ge 0$, on the other hand, then the spectrum of $ H^{\beta,a}$
  is asymptotically unbounded from below,
  \begin{equation*}
    \inf \spec { H^{\beta,a}}
%    = -\Err(a^{-2})  % only for beta>0
    \to -\infty \quad\text{as} \quad a \to 0.
  \end{equation*}
\end{lemma}
Note that although we know the limit spectrum as $a \to 0$ (see
\Lem{spec.delta'}), the resolvent convergence of \Thm{delta'} does not
necessarily imply the uniform boundedness from below of $H^{\beta,a}$
(see \Rem{sp.bdd2}).
\begin{proof} Let $\beta < 0$. Then an eigenfunction on the (original)
  edge $e$ has the form
  \begin{equation*}
    f_e(s)=
    \begin{cases}
      A \cosh (\kappa s) + B_e \sinh(\kappa s),& 0\le s \le a\\
      C_e \cosh (\kappa (1-s)),& a\le s \le 1.
    \end{cases}
  \end{equation*}
  for $\kappa>0$, the corresponding eigenvalue being
  $\lambda=-\kappa^2$. The Neumann condition $f_e'(1)=0$ at $s=1$ is
  automatically fulfilled, as well as the continuity at $s=0$ for the
  different edges $e$, since $f_e(0)=A$ is independent of $e$. The
  continuity in $s=a$ and the jump condition in the derivative lead to
  the system of equations
  \begin{gather*}
     A\cosh(\kappa a)  + B_e \sinh(\kappa a)
     - C_e \cosh(\kappa(1-a))
    =0\\
     -\frac 1a C_e \cosh \kappa(1-a)
%    -\frac 1a f_e(a)=f_e'(a+)-f_e'(a-)
    = \kappa\bigl(
      - A\sinh(\kappa a) - B_e \cosh(\kappa a)
      -C_e \sinh \kappa (1-a)
      \bigr)\\
     -\frac \beta{a^2} A
%    -\frac \beta{a^2} f(0)
%    =\sum_e f_e'(0)
    = \kappa \sum_e B_e.
  \end{gather*}
  With the permutational invariance in mind, let us first analyse the situation
  with symmetric coefficients, $A$, $B=B_e$, $C=C_e$. Then $\sum_e B_e
  = n B$ and the corresponding coefficient matrix for $A$, $B$ and $C$
  vanishes \emph{iff}
  \begin{equation*}
    \frac \beta {a^2}
    \bigl(\sinh(\kappa a) \cosh \kappa(1-a)
       - a \kappa \cosh \kappa \bigr)
     + n \kappa
    \bigl(
       \kappa a \sinh \kappa
       - \cosh(\kappa a) \cosh \kappa(1-a)
    \bigr)=0
  \end{equation*}
  leading to an eigenvalue $\lambda=-\kappa(a)^2$ of multiplicity one.
  It can be seen that $\kappa(a)$ is bounded, and that the above
  equation reduces to~\eqref{eq:kappa.beta} as $a \to \infty$.  The
  other eigenvalues can be obtained from $B$ and $C$ as follows: set
  $\Theta_n:= \e^{2 \pi \im/n}$.  Then for $k=1,\dots,n-1$, we have
  the coefficients $B_{e,k}=\Theta_n^{e \cdot k} B$ and
  $C_{e,k}=\Theta_n^{e \cdot k} C$, $e=1,\dots, n$. Since
  \begin{equation*}
    \sum_{e=1}^n B_{e,k} = B\sum_{e=0}^{n-1} \Theta_n^{e \cdot k} = 0
  \end{equation*}
  for $k=1,\dots, n-1$, we finally arrive at a
  coefficient matrix similar to the previous one, but with $n$
  replaced by zero. Consequently, if there were additional negative
  eigenvalues $\lambda=-\kappa(a)^2$, they would be of multiplicity
  $n-1$ and given by the relation
  \begin{equation*}
    \sinh(\kappa a) \cosh \kappa(1-a)
       - \kappa a \cosh \kappa
    =0.
  \end{equation*}
  But this equation has no solutions for $0<a\le 1$ and $\kappa>0$. We
  skip the proof of this fact here.

  For the second part, assume that $\beta \ge 0$. It is sufficient to
  calculate the Rayleigh quotient for the constant test function $f=\1
  \in \Sob G$ which yields
  \begin{equation*}
    \frac {\qf h^{\beta,a}(f)}{\normsqr f}
    =-\frac 1n \Bigl(\frac \beta {a^2} + \frac 1 a \Bigr)
  \end{equation*}
  being of order $\Err(a^{-2})$ if $\beta < 0$ and of order
  $\Err(a^{-1})$ if $\beta=0$, negative in both cases; recall that
  $n=\deg v$.
\end{proof}

Similarly, we expect the same behaviour for the operators on the
manifold.
\begin{lemma}
  \label{lem:lower.bd.mfd}
  If $\beta \ge 0$, then the spectrum of $H_\eps^\beta$ is
  asymptotically unbounded from below, i.e.,
  \begin{equation*}
    \inf \spec {H_\eps^\beta}
%    = -\Err(\eps^{-2\alpha})
    \to -\infty \quad\text{as} \quad \eps \to 0.
  \end{equation*}
\end{lemma}
\begin{proof}
%  (a) proof???
  Again, we plug the constant test function $u=\1$ into the Rayleigh
  quotient and obtain
  \begin{equation*}
    \frac {\qf h^\beta_\eps(u)}{\normsqr u}
    =-\frac {\beta \eps^{-2a} + \eps^{-a}}
          {n(1+\eps+\eps^\alpha) + \eps \vol X_{v_0}}
  \end{equation*}
  which obviously tends to $-\infty$ as $\eps \to 0$.
\end{proof}

\begin{remark}
  \label{rem:sp.bdd}
  As for a counterpart to the other claim in \Lem{lower.bd.gr}, the
  proof of the uniform boundedness from below as $\eps \to 0$ for
  $\beta < 0$ seems to need quite subtle estimates to compare the
  effect of the two competing potentials on $X_{\eps,v_0}$ and
  $X_{\eps,v_e}$ having strength proportional to
  $\abs{\beta}\eps^{-2\alpha}$ and $\eps^{-\alpha}$, respectively.
  Since the positive contribution $Q_{\eps,v_0}=\abs \beta
  \eps^{-1-2\alpha}$ is more singular than the negative contributions
  $Q_{\eps,v_e}=- \eps^{-1-\alpha}$, we expect that the threshold of
  the spectrum remains bounded as $\eps \to 0$.
\end{remark}

We can now prove our second main result. For the
$\delta'_\mathrm{s}$-coupling Hamiltonian $H_\beta$ and the
approximating operator $H_\eps^\beta$ defined in~\eqref{eq:h.beta}
and~\eqref{eq:h.beta.eps}, respectively, we make the following claim.
\begin{theorem}
  \label{thm:res.delta'}
  Assume that $0 < \alpha < 1/13$, then
  \begin{equation*}
    \bignorm{(H_\eps^\beta-\im)^{-1} J - J
      (H^\beta -\im)^{-1}}
    \to 0
  \end{equation*}
  as $\eps \to 0$.
\end{theorem}
\begin{proof}
  Denote by $H^{\beta,\eps} = H^{\beta,a_\eps}$ the $\eps$-depending
  intermediate Hamiltonian on the metric graph with
  $\delta$-potentials of strength depending on $\eps$ as defined
  before.  For the corresponding graph and manifold model, the lower
  bound to lengths depends now on $\eps$, specifically,
  $\ell_-=a_\eps=\eps^\alpha$.  Moreover, from the definition of the
  constants $C_{1/2} \le \wt C_{1/2}$ and $\eps_{1/2}$
  in~\eqref{eq:def.eps.eta} and from \Lem{closeness}, we conclude that
  \begin{equation*}
    \wt C_{1/2} = \wt C_{1/2}(\eps) =\Err(\eps^{-4\alpha}),
       \qquad
    \eps_{1/2}=\eps_{1/2}(\eps)=\Err(\eps^{3\alpha})
       \Und
    \delta=\delta_\eps = \Err(\eps^{(1-5\alpha)/2}).
  \end{equation*}
  Note that the dominant term in the closeness-error $\delta_\eps$
  (see~\eqref{eq:def.delta}) is the last one containing the potential.
  From \Thm{closeness} it follows now that
  \begin{equation*}
    \bignorm{(H_\eps^\beta-\im)^{-1} J - J
             ( H^{\beta,\eps}-\im)^{-1}}
    \le 10 \delta_\eps \max \{ \wt C_{1/2}(\eps), \sqrt 2\}
    = \Err(\eps^{(1-13\alpha)/2}).
  \end{equation*}
  so that \Thm{delta'} yields the sought conclusion. Note that the
  exponent of $\eps$ in $\delta_\eps \wt C_{1/2}(\eps)$ is
  $(1-5\alpha)/2 - 4\alpha= (1-13\alpha)/2>0$ provided $0 < \alpha <
  1/13$.
\end{proof}

We can now proceed and state similar results as in
\ThmS{res}{non-compact} for the $\delta'_\mathrm{s}$-approximation
by using arguments similar to those in~\cite[App.]{post:06}
or~\cite{post:pre08a}, where only non-negative operators were
considered (covering, as usual, operators bounded \emph{uniformly}
from below by a suitable shift). In our present situation, we can
only guarantee the resolvent convergence at \emph{non-real} points
like $z=\im$. Nevertheless, the arguments in~\cite[App.]{post:06}
or~\cite{post:pre08a} can be used to conclude the convergence of
suitable functions of operators as well as the convergence of the
dimension of spectral projections.

\begin{remark}
  \label{rem:sp.bdd2}
  Note that the asymptotic lower unboundedness of $H_\eps^\beta$ (and
  of the intermediate operator $H^{\beta,\eps}$) for $\beta \ge 0$
  described in \Lems{lower.bd.gr}{lower.bd.mfd} is not a contradiction
  to the fact that the limit operator $H^\beta$ is non-negative.  For
  example, the spectral convergence of \Thm{spec} holds only for
  \emph{compact} intervals $I \subset \R$.  In particular, $\spec {
    H^\beta} \cap I = \emptyset$ implies that
  \begin{equation*}
    \spec {H_\eps^\beta} \cap I = \emptyset
    \Und
    \spec {H^{\beta,\eps}} \cap I = \emptyset
  \end{equation*}
  provided $\eps>0$ is sufficiently small. This spectral convergence
  means that the negative spectral branches of $H_\eps^\beta$ all have
  to tend to $-\infty$.
\end{remark}

\subsection*{Acknowledgement}

O.P.\ enjoyed the hospitality in the Doppler Institute where a part
of the work was done. The research was supported by the Czech
Ministry of Education, Youth and Sports within the project
LC06002.

%----------------------------------------------------------------------
%\bibliographystyle{amsalpha}
%\bibliography{/home/post/Aktuell/BibTeX/literatur}
%----------------------------------------------------------------------
\newcommand{\etalchar}[1]{$^{#1}$}
\providecommand{\bysame}{\leavevmode\hbox to3em{\hrulefill}\thinspace}
\providecommand{\MR}{\relax\ifhmode\unskip\space\fi MR }
\renewcommand{\MR}[1]{}  % don't want MR-numbers in list

\end{document}